# *Ubah?*: Voting pattern in a sectarianized post-information-monopoly population
# I. predictions from a simple projection model

W. A. T. Wan Abdullah
*Physics Department,*
*Universiti Malaya,*
*50603 Kuala Lumpur,*
*Malaysia*
(wat@um.edu.my)

29 April 2013

**Abstract.** We predict the results of the upcoming elections in Malaysia using a projective model for a sectarianized post-information-monopoly population.

## Introduction

A central question in econophysics and sociophysics is how far can a collection of humans be described as a statistical system, like an ensemble of interacting atoms. Do the complex interaction between human agents force us to go beyond conventional statistical mechanics, or are they washed out in the limit of large agent numbers?

Some insight into this issue can be obtained by looking at large social phenomena like general elections. Here we predict the results of the upcoming elections in Malaysia (GE13, scheduled May 5$^{th}$ 2013, a few days in the future at the point of writing this) using a projective model for a sectarianized post-information-monopoly population. We await the real results in a few days to judge whether such models, with complexity washed out, can be acceptable.

## Geopolitics

Malaysia, with a population of about 27 million people, consisting of native Malays (and several others) and naturalized initially immigrant (about third generation now) ethic Chinese and Indians mostly, gained independence from British colonizers in 1957 and formed a federation of 11 (later 13) states plus several federal territories. Citizens vote for representatives to form the federal government as well as for those to form the respective state governments.

Since independence, the *Barisan Nasional* (BN) and its predecessor has been ruling, but now a viable competitor has arrived in the form of a coalition of several parties, called *Pakatan Rakyat* (PR). There are also some other small parties like BERJASA, SAPP, etc. Since independence, the ruling BN has more or less monopolized information through print and broadcast media monopoly, but recently this has been challenged by the advent of the internet and popular social media. This has moved support away from BN to PR, as can be deduced from the 2008 election results, when states in internet-dense regions fell to PR.

## Model

We make predictions using information on voter sizes and their ethnic fractions, and the majorities from 2008, which are freely available (see e.g. [1]). We only look at general statistical shifts in respective constituencies, largely ignoring locality and individual candidate personality effects.

It can be generally agreed that there would be a 5-10% swing in Malay votes towards PKR, judging from sentiments but dampened by misinformation effects. The actual value would depend on the degree of access to social media content of the various voting locations. We assume that the Chinese has already swung in 2008, as can be inferred from the voting pattern in 2008, and thus would not contribute any effective swing, and that the Indian and smaller communities also would contribute negligibly, given swing times voter size. We also assume that there would be roughly 70% voter turnout. As for the effects of new registered voters, we assume that the respective voter patterns would scale up.

It follows then that, if the Malay voter fraction for a constituency of voter size $N$ is $\alpha$, and the 2008 BN majority is $m$, then the needed percentage Malay swing for a PR win is

$$\sigma = 100m/(0.7\,\alpha\,N)$$

If for the particular federal constituency there was no contest in 2008, then we estimate the majority from the results for the corresponding state seats. We compared these values with the expected swing magnitude available constituency by constituency.

**Results**

The results are shown in the following table. It predicts a change of federal government to PR, and the win of 6 states to PR compared to 7 to BN (Sarawak state elections have been executed prior to this, with BN winning).

| | Number of seats predicted | | | |
|---|---|---|---|---|
| | Government | BN | PR | others |
| Federal | PR | 94 | 127 | 1 |
| Perlis state | BN | 11 | 4 | |
| Kedah state | PR | 3 | 33 | |
| Kelantan state | PR | 3 | 42 | |
| Terengganu state | PR | 14 | 18 | |
| Pulau Pinang state | PR | 3 | 37 | |
| Perak state | PR | 13 | 46 | |
| Pahang state | BN | 29 | 13 | |
| Selangor state | PR | 8 | 48 | |
| Negeri Sembilan state | BN | 19 | 17 | |
| Melaka state | BN | 20 | 8 | |
| Johor state | BN | 45 | 11 | |
| Sabah state | BN | 51 | 8 | 1 |

The constituent-by-constituent predictions are given in the Appendix.

**Conclusions**

We have predicted the results of the upcoming elections in Malaysia using a projective model for a sectarianized post-information-monopoly population. We await the real results to see if the model and its basis are acceptable.

**Reference**

[1] *Berita Harian*, 21 April 2013.

**Appendix**

| Federal Government | | |
|---|---|---|
| Area | BN (x) / PR (o)/ others (-) | Notes |
| P001 | x | |
| P002 | x | |
| P003 | o | |
| P004 | x | |
| P005 | o | |
| P006 | x | |
| P007 | o | |
| P008 | o | |
| P009 | o | BERJASA may pull some Malay votes from PR |
| P010 | o | |
| P011 | o | |
| P012 | o | |
| P013 | o | |
| P014 | o | |
| P015 | o | |
| P016 | o | |
| P017 | o | |
| P018 | o | |
| P019 | o | |
| P020 | o | |
| P021 | o | |
| P022 | o | |
| P023 | o | |
| P024 | o | |
| P025 | o | |
| P026 | o | |
| P027 | o | |
| P028 | o | |
| P029 | o | |
| P030 | x | |
| P031 | o | |
| P032 | x | |
| P033 | x | |
| P034 | x | |
| P035 | o | |
| P036 | o | |
| P037 | o | |
| P038 | x | |
| P039 | o | |
| P040 | x | |
| P041 | x | |
| P042 | o | |
| P043 | o | |
| P044 | o | |
| Area | BN (x) / PR (o)/ others (-) | Notes |
| P045 | ○ | |
| P046 | ○ | |
| P047 | ○ | |
| P048 | ○ | |
| P049 | ○ | |
| P050 | ○ | |
| P051 | ○ | |
| P052 | ○ | |
| P053 | ○ | |
| P054 | × | |
| P055 | × | |
| P056 | ○ | |
| P057 | ○ | |
| P058 | ○ | |
| P059 | ○ | |
| P060 | ○ | |
| P061 | ○ | |
| P062 | ○ | |
| P063 | ○ | |
| P064 | ○ | |
| P065 | ○ | |
| P066 | ○ | |
| P067 | ○ | |
| P068 | ○ | |
| P069 | ○ | |
| P070 | × | |
| P071 | ○ | |
| P072 | × | |
| P073 | ○ | |
| P074 | ○ | |
| P075 | × | |
| P076 | ○ | |
| P077 | × | |
| P078 | × | |
| P079 | × | |
| P080 | ○ | |
| P081 | ○ | |
| P082 | ○ | |
| P083 | ○ | |
| P084 | × | |
| P085 | × | |
| P086 | × | |
| P087 | × | |
| P088 | ○ | |
| P089 | × | |
| P090 | × | |
| P091 | × | |
| P092 | ○ | |
| P093 | × | |
| P094 | ○ | |
| P095 | ○ | |
| P096 | ○ | |
| P097 | ○ | BERJASA may pull some Malay votes from PR |
| P098 | ○ | |
| P099 | ○ | |
| P100 | ○ | |
| P101 | ○ | |
| P102 | ○ | |
| P103 | ○ | |
| P104 | ○ | |
| P105 | ○ | |
| P106 | ○ | |
| P107 | ○ | |
| P108 | ○ | |
| P109 | ○ | BERJASA may pull some Malay votes from PR |
| P110 | ○ | |
| P111 | ○ | |
| P112 | ○ | |
| P113 | ○ | |
| P114 | ○ | |
| P115 | ○ | BERJASA may pull some Malay votes from PR |
| P116 | ○ | |
| P117 | ○ | |
| P118 | x | |
| P119 | ○ | |
| P120 | ○ | |
| P121 | ○ | |
| P122 | ○ | |
| P123 | ○ | |
| P124 | ○ | |
| P125 | ○ | |
| P126 | x | |
| P127 | x | |
| P128 | ○ | BERJASA may pull some Malay votes from PR |
| P129 | x | |
| P130 | ○ | |
| P131 | ○ | |
| P132 | ○ | |
| P133 | x | |
| P134 | x | |
| P135 | x | |
| P136 | x | |
| P137 | ○ | |
| P138 | ○ | |
| P139 | x | |
| P140 | ○ | Close |
| P141 | x | |
| P142 | x | |
| P143 | x | |
| P144 | x | |
| P145 | o | |
| P146 | o | Close |
| P147 | x | |
| P148 | x | |
| P149 | x | |
| P150 | x | |
| P151 | x | |
| P152 | x | |
| P153 | x | |
| P154 | x | |
| P155 | x | |
| P156 | x | |
| P157 | x | |
| P158 | x | |
| P159 | x | |
| P160 | x | |
| P161 | x | |
| P162 | x | |
| P163 | x | |
| P164 | x | |
| P165 | x | |
| P166 | x | |
| P167 | x | |
| P168 | o | |
| P169 | o | |
| P170 | o | |
| P171 | - | SAPP was big winner in last elections |
| P172 | o | |
| P173 | x | |
| P174 | o | |
| P175 | x | |
| P176 | x | |
| P177 | o | Last winner was in BN now standing in PR |
| P178 | x | |
| P179 | x | |
| P180 | o | Sectarian votes may split between BN, STAR |
| P181 | x | |
| P182 | x | |
| P183 | x | |
| P184 | x | |
| P185 | X | |
| P186 | O | |
| P187 | X | |
| P188 | X | |
| P189 | X | |
| P190 | O | |
| P191 | O | |
| P192 | X | |
| P193 | X | |
| P194 | X | |
| P195 | O | |
| P196 | O | |
| P197 | X | |
| P198 | O | |
| P199 | X | |
| P200 | X | |
| P201 | X | |
| P202 | O | Big swing assumed |
| P203 | O | |
| P204 | X | |
| P205 | X | |
| P206 | X | |
| P207 | X | |
| P208 | O | |
| P209 | X | |
| P210 | X | |
| P211 | O | |
| P212 | O | |
| P213 | X | |
| P214 | X | |
| P215 | X | |
| P216 | O | |
| P217 | X | |
| P218 | X | |
| P219 | O | |
| P220 | X | |
| P221 | O | |
| P222 | X | |

| Perlis State Government | | |
|---|---|---|
| Area | BN (x) / PR (o)/ others (-) | Notes |
| N01 | x | |
| N02 | x | |
| N03 | x | |
| N04 | x | |
| N05 | o | |
| N06 | x | |
| N07 | x | |
| N08 | x | |
| N09 | x | |
| N10 | x | |
| N11 | x | |
| N12 | x | |
| N13 | o | |
| N14 | o | |
| N15 | o | |

| Kedah State Government | | |
|---|---|---|
| Area | BN (x) / PR (o)/ others (-) | Notes |
| N01 | o | |
| N02 | x | |
| N03 | o | |
| N04 | o | |
| N05 | x | |
| N06 | o | |
| N07 | o | |
| N08 | x | |
| N09 | o | |
| N10 | o | |
| N11 | o | BERJASA may pull some Malay votes from PR |
| N12 | o | |
| N13 | o | |
| N14 | o | |
| N15 | o | |
| N16 | o | |
| N17 | o | |
| N18 | o | |
| N19 | o | |
| N20 | o | |
| N21 | o | |
| N22 | o | |
| N23 | o | |
| N24 | o | |
| N25 | o | |
| N26 | o | |
| N27 | o | |
| N28 | o | |
| N29 | o | |
| N30 | o | |
| N31 | o | |
| N32 | o | |
| N33 | o | |
| N34 | o | |
| N35 | o | BERJASA may pull some Malay votes from PR |
| N36 | o | |

| Kelantan State Government | | |
|---|---|---|
| Area | BN (x) / PR (o)/ others (-) | Notes |
| N01 | o | |
| N02 | o | |
| N03 | o | |
| N04 | o | |
| N05 | o | |

| N06 | ○ | |
| N07 | ○ | |
| N08 | ○ | |
| N09 | ○ | |
| N10 | ○ | |
| N11 | ○ | |
| N12 | ○ | |
| N13 | ○ | |
| N14 | ○ | |
| N15 | ○ | |
| N16 | ○ | |
| N17 | ○ | |
| N18 | ○ | |
| N19 | ○ | |
| N20 | ○ | |
| N21 | ○ | |
| N22 | ○ | |
| N23 | ○ | |
| N24 | ○ | |
| N25 | ○ | |
| N26 | ○ | |
| N27 | ○ | |
| N28 | ○ | |
| N29 | ○ | |
| N30 | ○ | |
| N31 | ○ | |
| N32 | ○ | |
| N33 | ○ | |
| N34 | ○ | |
| N35 | ○ | |
| N36 | ○ | |
| N37 | ○ | |
| N38 | x | |
| N39 | ○ | |
| N40 | ○ | |
| N41 | ○ | |
| N42 | ○ | |
| N43 | x | |
| N44 | x | |
| N45 | ○ | |

| Terengganu State Government | | |
|---|---|---|
| Area | BN (x) / PR (o)/ others (-) | Notes |
| N01 | x | |
| N02 | x | |
| N03 | x | |
| N04 | x | |
| N05 | ○ | |

| N06 | x | |
| N07 | x | |
| N08 | ○ | |
| N09 | ○ | |
| N10 | ○ | |
| N11 | x | |
| N12 | ○ | |
| N13 | ○ | |
| N14 | ○ | Malay sectarian votes help PR |
| N15 | ○ | |
| N16 | ○ | |
| N17 | ○ | |
| N18 | ○ | |
| N19 | ○ | |
| N20 | ○ | |
| N21 | x | |
| N22 | 0 | |
| N23 | x | |
| N24 | x | |
| N25 | x | |
| N26 | ○ | |
| N27 | ○ | |
| N28 | ○ | |
| N29 | x | |
| N30 | x | |
| N31 | ○ | |
| N32 | x | |

| Pulau Pinang State Government | | |
|---|---|---|
| Area | BN (x) / PR (o)/ others (-) | Notes |
| N01 | ○ | |
| N02 | x | |
| N03 | x | |
| N04 | x | |
| N05 | ○ | |
| N06 | ○ | |
| N07 | ○ | |
| N08 | ○ | |
| N09 | ○ | |
| N10 | ○ | |
| N11 | ○ | |
| N12 | ○ | |
| N13 | ○ | |
| N14 | ○ | One PR cand. gives way, but may lose votes |
| N15 | ○ | |
| N16 | ○ | |
| N17 | ○ | |
| N18 | ○ | |

| Area | BN (x) / PR (o)/ others (-) | Notes |
|---|---|---|
| N19 | o | |
| N20 | o | |
| N21 | o | |
| N22 | o | |
| N23 | o | |
| N24 | o | |
| N25 | o | |
| N26 | o | |
| N27 | o | |
| N28 | o | |
| N29 | o | |
| N30 | o | |
| N31 | o | |
| N32 | o | |
| N33 | o | |
| N34 | o | |
| N35 | o | |
| N36 | o | |
| N37 | o | |
| N38 | o | |
| N39 | o | |
| N40 | o | |

| Perak State Government | | |
|---|---|---|
| Area | BN (x) / PR (o)/ others (-) | Notes |
| N01 | x | |
| N02 | x | |
| N03 | x | |
| N04 | x | |
| N05 | o | |
| N06 | o | |
| N07 | x | |
| N08 | o | |
| N09 | o | |
| N10 | o | |
| N11 | o | |
| N12 | o | |
| N13 | o | BERJASA may pull some Malay votes from PR |
| N14 | o | |
| N15 | o | |
| N16 | o | |
| N17 | o | |
| N18 | o | |
| N19 | o | |
| N20 | o | |
| N21 | o | |
| N22 | o | |
| N23 | o | |

| Area | BN (x) / PR (o)/ others (-) | Notes |
|---|---|---|
| N24 | x | |
| N25 | o | |
| N26 | o | |
| N27 | o | |
| N28 | o | |
| N29 | o | |
| N30 | o | |
| N31 | o | |
| N32 | o | |
| N33 | o | |
| N34 | o | |
| N35 | o | |
| N36 | o | |
| N37 | o | |
| N38 | x | |
| N39 | x | |
| N40 | o | |
| N41 | o | |
| N42 | x | |
| N43 | o | |
| N44 | o | |
| N45 | o | |
| N46 | x | |
| N47 | x | |
| N48 | o | |
| N49 | o | |
| N50 | o | |
| N51 | o | |
| N52 | x | |
| N53 | o | |
| N54 | o | |
| N55 | o | |
| N56 | o | |
| N57 | o | |
| N58 | x | |
| N59 | o | |

| Pahang State Government | | |
|---|---|---|
| Area | BN (x) / PR (o)/ others (-) | Notes |
| N01 | x | |
| N02 | x | |
| N03 | x | |
| N04 | x | |
| N05 | x | |
| N06 | x | |
| N07 | o | |
| N08 | x | |
| N09 | x | |

| Area | BN (x) / PR (o)/ others (-) | Notes |
|---|---|---|
| N10 | o | |
| N11 | x | |
| N12 | o | |
| N13 | x | |
| N14 | o | |
| N15 | o | |
| N16 | o | |
| N17 | x | |
| N18 | x | |
| N19 | o | |
| N20 | x | |
| N21 | x | |
| N22 | x | |
| N23 | x | |
| N24 | x | |
| N25 | x | |
| N26 | x | |
| N27 | x | |
| N28 | x | |
| N29 | o | |
| N30 | x | |
| N31 | x | |
| N32 | o | |
| N33 | o | |
| N34 | o | |
| N35 | o | |
| N36 | x | |
| N37 | x | |
| N38 | o | |
| N39 | x | |
| N40 | x | |
| N41 | x | |
| N42 | x | |

| Selangor State Government | | |
|---|---|---|
| Area | BN (x) / PR (o)/ others (-) | Notes |
| N01 | o | |
| N02 | o | |
| N03 | x | |
| N04 | o | |
| N05 | x | |
| N06 | o | |
| N07 | o | |
| N08 | x | |
| N09 | o | |
| N10 | o | |
| N11 | o | |
| N12 | o | |

| Area | BN (x) / PR (o)/ others (-) | Notes |
|---|---|---|
| N13 | o | |
| N14 | o | |
| N15 | o | |
| N16 | o | |
| N17 | o | |
| N18 | o | |
| N19 | o | |
| N20 | o | |
| N21 | o | |
| N22 | o | |
| N23 | o | |
| N24 | o | |
| N25 | o | |
| N26 | o | |
| N27 | o | |
| N28 | o | |
| N29 | o | |
| N30 | o | |
| N31 | o | |
| N32 | o | |
| N33 | o | |
| N34 | o | |
| N35 | o | |
| N36 | o | |
| N37 | o | |
| N38 | o | |
| N39 | x | Two PR candidates. Votes diluted. |
| N40 | o | |
| N41 | x | |
| N42 | o | |
| N43 | o | |
| N44 | o | |
| N45 | o | |
| N46 | o | |
| N47 | o | |
| N48 | o | |
| N49 | o | |
| N50 | o | |
| N51 | o | |
| N52 | o | |
| N53 | o | |
| N54 | x | |
| N55 | x | |
| N56 | x | |

| Negeri Sembilan State Government | | |
|---|---|---|
| Area | BN (x) / PR (o)/ others (-) | Notes |
| N01 | X | |
| N02 | X | |

| Area | BN (x) / PR (o)/ others (-) | Notes |
|---|---|---|
| N03 | X | |
| N04 | X | Close |
| N05 | X | |
| N06 | X | |
| N07 | X | |
| N08 | O | |
| N09 | O | |
| N10 | O | |
| N11 | O | |
| N12 | O | |
| N13 | O | |
| N14 | O | |
| N15 | X | |
| N16 | X | |
| N17 | X | |
| N18 | X | |
| N19 | X | |
| N20 | O | |
| N21 | O | |
| N22 | O | |
| N23 | O | |
| N24 | O | |
| N25 | O | |
| N26 | X | |
| N27 | X | |
| N28 | X | |
| N29 | O | |
| N30 | O | |
| N31 | X | |
| N32 | X | |
| N33 | O | |
| N34 | X | |
| N35 | X | |
| N36 | O | |

| Melaka State Government | | |
|---|---|---|
| Area | BN (x) / PR (o)/ others (-) | Notes |
| N01 | x | |
| N02 | x | |
| N03 | x | |
| N04 | x | |
| N05 | x | |
| N06 | x | |
| N07 | x | |
| N08 | x | |
| N09 | x | |
| N10 | x | |
| N11 | x | |

| Area | BN (x) / PR (o)/ others (-) | Notes |
|---|---|---|
| N12 | x | |
| N13 | o | Close |
| N14 | o | |
| N15 | o | |
| N16 | o | |
| N17 | x | |
| N18 | x | |
| N19 | o | |
| N20 | o | |
| N21 | o | |
| N22 | o | |
| N23 | x | |
| N24 | x | |
| N25 | x | |
| N26 | x | |
| N27 | x | |
| N28 | x | |

| Johor State Government | | |
|---|---|---|
| Area | BN (x) / PR (o)/ others (-) | Notes |
| N01 | x | |
| N02 | o | Close |
| N03 | x | |
| N04 | x | |
| N05 | x | |
| N06 | x | |
| N07 | x | |
| N08 | x | |
| N09 | x | |
| N10 | o | |
| N11 | x | |
| N12 | o | |
| N13 | o | |
| N14 | x | |
| N15 | o | |
| N16 | o | Close |
| N17 | x | |
| N18 | x | |
| N19 | x | |
| N20 | x | |
| N21 | x | |
| N22 | x | |
| N23 | o | |
| N24 | x | |
| N25 | x | |
| N26 | x | |
| N27 | x | |
| N28 | o | |
| N29 | x | |
| N30 | x | |
| N31 | x | |
| N32 | x | |
| N33 | x | |
| N34 | x | |
| N35 | x | |
| N36 | x | |
| N37 | x | |
| N38 | x | |
| N39 | x | |
| N40 | x | |
| N41 | x | |
| N42 | x | |
| N43 | o | Close |
| N44 | x | |
| N45 | x | |
| N46 | x | |
| N47 | x | |
| N48 | o | |
| N49 | x | |
| N50 | x | |
| N51 | x | |
| N52 | o | |
| N53 | x | |
| N54 | x | |
| N55 | x | |
| N56 | x | |

| Sabah State Government | | |
|---|---|---|
| Area | BN (x) / PR (o)/ others (-) | Notes |
| N01 | x | |
| N02 | x | |
| N03 | x | |
| N04 | x | |
| N05 | x | |
| N06 | x | |
| N07 | x | |
| N08 | x | |
| N09 | x | |
| N10 | x | |
| N11 | x | |
| N12 | x | |
| N13 | - | SAPP cand. won big Federal last elections |
| N14 | o | |
| N15 | o | |
| N16 | o | |
| N17 | x | |
| N18 | × | |
| N19 | ○ | |
| N20 | × | |
| N21 | × | |
| N22 | × | |
| N23 | × | |
| N24 | × | |
| N25 | × | |
| N26 | ○ | |
| N27 | × | |
| N28 | × | |
| N29 | × | |
| N30 | × | |
| N31 | × | |
| N32 | × | |
| N33 | ○ | |
| N34 | × | |
| N35 | × | |
| N36 | × | |
| N37 | × | |
| N38 | × | |
| N39 | × | |
| N40 | × | |
| N41 | × | |
| N42 | × | |
| N43 | × | |
| N44 | × | |
| N45 | × | |
| N46 | × | |
| N47 | × | |
| N48 | × | |
| N49 | × | |
| N50 | × | |
| N51 | × | |
| N52 | × | |
| N53 | × | |
| N54 | × | |
| N55 | × | |
| N56 | × | |
| N57 | ○ | |
| N58 | ○ | |
| N59 | × | |
| N60 | × | |